\documentclass[twocolumn,preprintnumbers,amsmath,amssymb,prb]{revtex4}
\usepackage{dcolumn}% Align table columns on decimal point
\usepackage{bm}% bold math
\usepackage[dvips]{graphicx}
\begin{document}
\title{Flux-flow induced giant magnetoresistance in all-amorphous superconductor-ferromagnet hybrids}
\author{C. Bell, S. Tur\c{s}ucu and J. Aarts}
\affiliation{Magnetic and Superconducting Materials Group, \\ Kamerlingh Onnes
Laboratorium, Universiteit Leiden, Leiden, The Netherlands}
\date{\today}
\begin{abstract}
We present magnetoresistance measurements on all-amorphous ferromagnet (F) /
superconductor (S) heterostructures. The F/S/F trilayers show large
magnetoresistance peaks in a small field range around the coercive field of the F
layers, at temperatures within and below the superconducting transition. This is
interpreted as flux flow of weakly pinned vortices induced by the stray field of
Bloch magnetic domains in the F layers. Bilayers show much smaller effects, implying
that the Bloch walls of the F-layers in the trilayer line up and focus the stray
fields. The data are used to discuss the expected minimum F-layer thickness needed
to nucleate vortices.
\end{abstract}
%\pacs{}
\maketitle
\section{Introduction}
There are a number of phenomena currently under investigation which involve the
combinations of superconductors (S) and ferromagnets (F). In S/F/S configurations,
the superconductors can be coupled through the ferromagnetic layer, which may lead
to so-called $\pi$-junctions.\cite{buzdin05} In F/S/F configurations, the
superconducting transition temperature (T$_c$) depends on the relative orientation
of the magnetization in two F layers,\cite{tagirov99,baladie03,fominov03,gu02} which
constitutes the so-called superconducting spin switch. In researching these
phenomena, the question of the influence of domain structures in the F layers is
often ignored. Domain walls can have various effects. For instance,
superconductivity can be enhanced by domain walls, through two different mechanisms.
One is that Cooper pairs sample inhomogeneous exchange interactions in the wall, or
the different directions of the magnetization on the two sides of the wall, and
experience less pair breaking. This was observed in bilayers of Nb and Permalloy
(Py) \cite{rusanov06} and also in Nb/Co.\cite{kinsey01} The other mechanism comes
about in ferromagnets with a preferred magnetization direction perpendicular to the
plane of a superconducting film. Now, the presence of a domain wall can lead to a
local reduction of the amount of flux through the superconductor, and therefore to
less suppression (or relative enhancement) of superconductivity. This was
demonstrated on S/F bilayers involving Nb and a ferromagnetic garnet
(BaFe$_{12}$O$_{19}$)\cite{yang04}, and on S/I/F bilayers (with I an insulating
barrier) and F/S/F trilayers with Pb or Nb as the S layer combined with
perpendicularly magnetized Co/Pd multilayers\cite{lange03,gillijns05}.

A different situation occurs when the magnetization of the ferromagnet is in-plane
and the magnetization in the domain wall rotates out of the plane (so-called Bloch
walls). This can influence the superconductivity in the S layer if the flux from the
wall creates vortices. Observations on Nb/CuNi bilayers were interpreted in this way
\cite{ryazanov03}, but otherwise the problem has received little attention
experimentally. Recently, the conditions for vortex formation were discussed
theoretically (see Ref. \cite{burmistrov05} and references therein). In this work we
present data from an experimental system well suited to observe the effects of
vortices in transport measurements, consisting of a combination of an amorphous
ferromagnet (a-Gd$_{19}$Ni$_{81}$, referred to as GdNi) and an amorphous
superconductor (a-Mo$_{2.7}$Ge, called MoGe). Because of the amorphous nature of the
materials, the magnet has an extremely low switching or coercive field $H_c$,
corresponding to an applied flux density of less than 1~mT, while the superconductor
has very weak vortex pinning properties. Also, the magnetic material has a
relatively high magnetization (due to the Gd atoms), which facilitates vortex
formation. In F/S/F trilayers, we show the occurrence of extremely sharp resistance
spikes when varying the magnetic field around $H_c$ at temperatures near the base of
the superconducting transition, which we interpret as due to flux flow. Furthermore,
we find that the effect is much weaker in F/S bilayers, presumably because domain
walls in both F-layers tend to line up, thereby focussing the flux coming out of the
walls. Such a coupling of the domain walls makes the F/S/F trilayer case different
from the case of F/S bilayers. After the presentation of the data, we apply the
model developed in Ref.~\cite{burmistrov05} to argue that vortices can be created in
the MoGe layer by the flux from the domain walls, and we discuss the requirements of
vortex formation in various other S/F systems.

\section{Single film characteristics}
Our samples are grown on (100) oxidized Si by r.f. sputtering at room temperature,
in a vacuum system with a base pressure $<$ 2$\times$10$^{-6}$ mbar. Deposition
rates were of the order of $\sim 7.5$~nm/min for the GdNi and $\sim 8.5$~nm/min for
the MoGe, as calibrated from low angle X-ray reflectivity. The compositions were
found using Rutherford Backscattering, and the amorphous nature of the films (i.e.
the absence of crystallinity) was checked by X-ray diffraction. The bulk
superconducting transition temperature $T_c$ of our MoGe films is about 5.5~K, and
such films show weak vortex pinning properties as reported on similar material grown
in the same system \cite{plourde02,baarle03}. Another particular property of
amorphous superconductors is that the very small mean free path (also reflected in a
large specific resistance of typically about $200\times 10^{-8} \Omega$m) leads to a
large zero-temperature London (magnetic) penetration depth $\lambda_L(0)$, of order
0.7~$\mu$m. The zero-temperature coherence length $\xi(0)$ of these films is small,
around 5~nm. The numbers result in a quite small value for the zero-temperature
lower critical field $H_{c1}(0)$ of typically $1.3\times 10^3$ A/m (corresponding to
1.6~mT), but in a very large value for the zero-temperature upper critical field
$H_{c2}(0)$ of $\sim$ 13~T.

Amorphous GdNi belongs to a general class of ferromagnets combining a rare earth
element and a transition metal element, which both carry a moment on their own
subnetwork in the material. The amorphous state leads to a spatial distribution of
the relative directions of the magnetic moments of both networks. If a net moment
exists, the state is called sperimagnetic \cite{moorjani84}. The moments of the two
networks are coupled antiferromagnetically, and since the temperature dependence of
the magnetization is different for both, there may exist a so-called compensation
temperature $T_{comp}$ where the two magnetizations cancel. The case of
Gd$_{1-x}$Ni$_x$ is a special one. According to the literature
\cite{moorjani84,asomoza79}, the Ni atom does not possess a magnetic moment below a
critical concentration $x_c \approx$ 0.8 while the Gd atoms have their full S-state
moment of about 7~$\mu_B$, which order ferromagetically. With increasing $x$, for $x
< x_c$, the dilution effect of the non-magnetic Ni on the Gd matrix leads to a
decrease of the magnetic ordering temperature $T_c$ down to 40~K at $x$ = $x_c
\approx 0.8$. Above $x_c$, a small moment appears on the Ni, antiparallel to the Gd
moment, leading to a simple ferrimagnetic state. Obviously, a compensation
temperature can only occur for $ x > x_c$.
\begin{figure}[h]
%Fig. 1
\includegraphics[width=8.5cm]{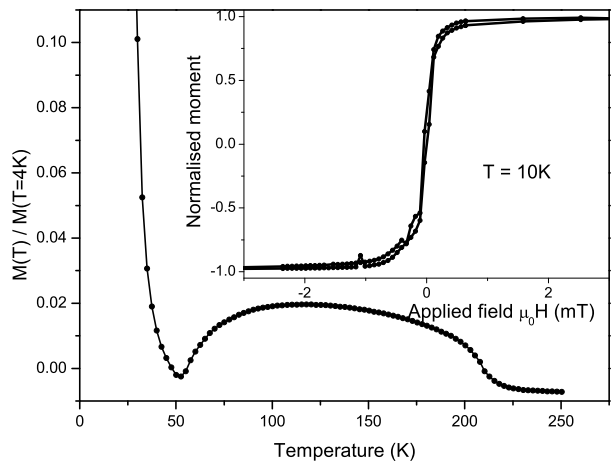}
\caption{\label{fig:MGdNi}Magnetic moment $M$, normalised to its value at 4~K, vs.
temperature $T$ for a 20~nm GdNi film in an applied field of $\mu_0 H = 3$ mT.
Inset: $M$ vs. applied field $\mu_0 H$ for the same film at $T$ = 10 K.}
\end{figure}

For our material with $x$ = 0.81, we find a saturation magnetization, measured at
10~K on a film of a 75~nm, of $7.8 \pm 0.2 \times 10^5$ A/m. Assuming the full Gd
moment of 7~$\mu_B$, this yields a small Ni moment of -0.02~$\mu_B$, in good
agreement with the earlier data \cite{note-density}. Fig~\ref{fig:MGdNi} shows the
temperature dependence of the magnetization (normalized to the value at 4~K) of a
20~nm film, measured in an applied field of 3~mT. Clearly visible is that the
magnetization already develops around 220~K, but dips again to 0 \cite{note-diamag}
at $T_{comp}$~= 50~K. These value are reduced slightly for the thinner films used in
the rest of this work, but the the alloy evidently orders at a much higher
temperature than previously reported. It seems quite likely that $T_c$ was mistaken
for $T_{comp}$, possibly because of a too high applied field. The inset of
Fig.~\ref{fig:MGdNi} shows the field dependence of the magnetization, taken at 10~K.
The coercive field $H_c$ is very small in this case, less than 0.1~mT, which is a
consequence of the absence of anisotropy in the Gd S-state, and the lack of grain
boundaries which hinder domain wall motion.

All samples were patterned with e-beam lithography and broad beam Ar ion milling to
100 $\mu$m wide wires for a standard four point measurement with 1~mm between
voltage contacts. The trilayer samples for which data is presented are of the form
GdNi(x)/MoGe(y)/GdNi(z) with x, y, z all in nm, and the first of these grown on the
SiO. All of the data presented are measured with the magnetic field applied in-plane
and (anti-)parallel to the current to within a few degrees error (no precise
alignment procedure was undertaken), and with a constant current of $\pm 100~\mu$A
unless otherwise stated.

\section{Results}
\begin{figure}[b]
%Fig.2
\includegraphics[width=8.5cm]{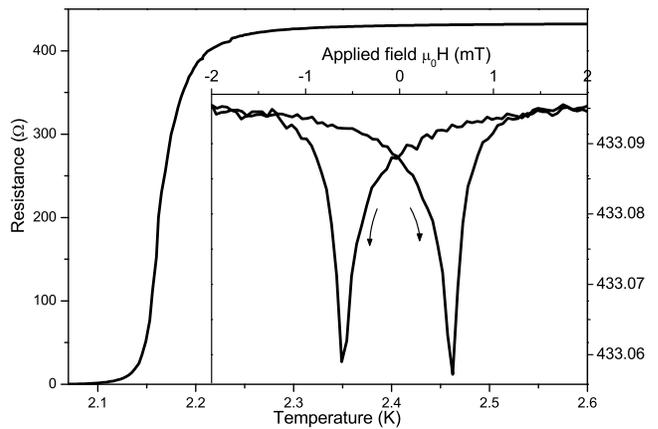}
\caption{\label{fig:RTtri}Superconducting resistive transition for a patterned wire
in a GdNi(11)/MoGe(21)/GdNi(11) trilayer. Inset: Magnetoresistance for $H \parallel
I$ at 3.75 K. The arrows denote the direction of the field sweep.}
\end{figure}
\begin{figure}[h]
%Fig.3
\includegraphics[width=8.5cm]{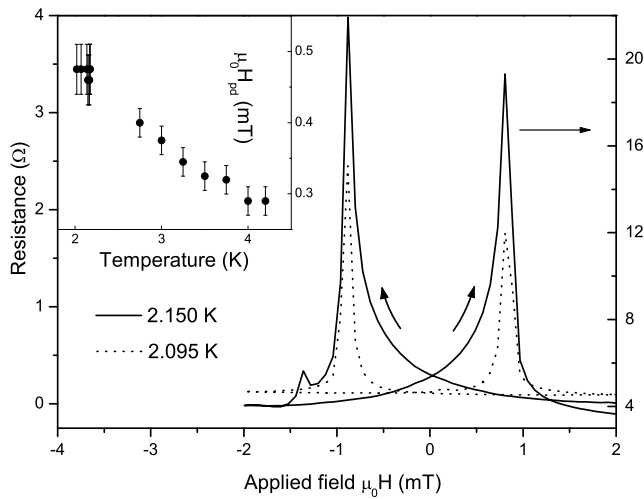}
\caption{\label{paperRvsH}Resistance vs. applied field of a
GdNi(11)/MoGe(21)/GdNi(11) trilayer for two temperatures within the superconducting
transition as indicated. The arrows denote the direction of the field sweep. Inset:
Values of the peak / dip field $H_{pd}$ in $R(H)$ vs. temperature, for a related
sample with slightly lower $H_c$.}
\end{figure}

Fig.~\ref{fig:RTtri} shows the superconducting transition of a sample
GdNi(11)/MoGe(21)/GdNi(11), with a midpoint at 2.16~K and a width (10~\% - 90~\%) of
60~mK. The reduced $T_c$ indicates a significant proximity effect from the F layers.
The inset shows the behavior of the resistance $R$ versus applied field $H$ for $H
\parallel$ current $I$ at 3.75~K (above the transition). Small dips are visible
around the switching field of the F layers at 0.5~mT, which is the conventional
anisotropic magnetoresistance with a magnitude $\Delta R / R_{max}$ of the order of
$9 \times 10^{-5}$. We observe a single peak in each quadrant, indicating that the
two layers switch at the same field. Fig.~\ref{paperRvsH} shows $R(H)$ at 2.15~K and
at 2.095~K, at the base of the transition. Sweeping the field now leads to strong
resistance peaks with $\Delta R$ several percent of the normal state resistance. For
a related sample (with slightly lower $H_c$) we also plot the peak (dip) position
$H_{pd}$ of the $R(H)$ measurements through the superconducting transition (inset
Fig.~\ref{paperRvsH}). It is clear that $H_c$ increases smoothly with $T$, and that
the large magnetoresistance peaks in the transition are associated with the domain
state of the F layers at $H_c$.
The maximum $\Delta R$ for this sample was $\sim 26$ $\Omega$, representing a change
of 6 \% of the total normal state resistance of the whole trilayer; however $\Delta
R$ decreases with decreasing T and with this measurement current of 100~$\mu A$ the
voltage in the peaks passes below a 1$\mu$V criterion at 2.03~K.

\begin{figure}[t]
%Fig.4
\includegraphics[width=8.5cm]{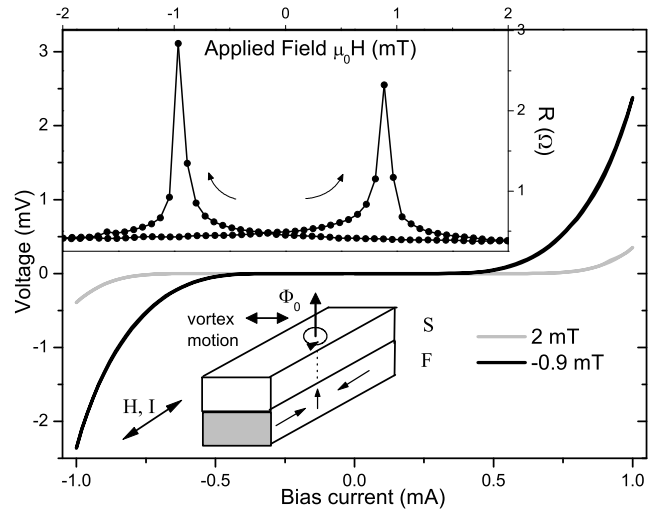}
\caption{\label{1_5mins_even_IV_1_8K}Current-voltage characteristics at T = 1.8 K at
an applied field $-H_c$ (drawn line) and slightly above $+H_c$ (dotted line) for the
GdNi(11)/MoGe(21)/GdNi(11) sample. Inset: $R(H)$ when biasing at 1 mA, above the
depinning current. The arrows denote the direction of the field sweep. Also shown is
a sketch of the sample configuration, with the directions of applied field, current,
and flux inside the F-layer (including the domain wall) as indicated. }
\end{figure}

Since the resistance shows a peak in the the domain state of the F layers, we cannot
interpret the data in terms of mechanisms which yield a decreased resistance
(enhanced superconductivity) when the relative magnetization directions in the two F
layers deviate from parallel. In principle, quasiparticle trapping could provide a
mechanism for increased resistance\cite{rusanov06}. It would then have to be argued
that the domain state locally leads to antiparallel configurations between the two F
layers, as was actually found in the case of F/S/F trilayers involving
(La,Ca)MnO$_3$ and YBa$_2$Cu$_3$O$_7$.\cite{pena05} However, in view of the weak
pinning properties of the superconductor, another possibility is flux flow
resistivity associated with the motion of vortices formed spontaneously above Bloch
domain walls. The sample is in the force-free configuration ($H \parallel I$) for
the applied field, but induced vortices pointing out of the plane of the sample will
feel a Lorenz force due to the applied current, which can cause vortex motion across
the width of the wire. To clarify this, we measured current (I) - voltage (V)
characteristics on the same sample GdNi(11)/MoGe(21)/GdNi(11) at 1.8~K, well below
the transition, and with the field either at -$H_c$ or sligtly above +$H_c$. They
are shown in Fig.\ref{1_5mins_even_IV_1_8K}, which also shows a sketch of the sample
configuration, with the directions of applied field, current, and flux inside the
F-layer (including the domain wall). At this temperature we find a true
supercurrent, and a gradual onset of voltage. Using a 1$\mu$V criterion, the
critical currents in the low and high resistive states taken from
Fig.\ref{1_5mins_even_IV_1_8K} are $\sim$ 640 and 340 $\mu$A respectively.
Resistance peaks at this temperature therefore can still be seen, as long as the
bias current is large enough to depin the vortices. This is shown in the inset of
Fig.\ref{1_5mins_even_IV_1_8K}, where the voltage (resistance) was taken at a bias
current of 1~mA.

\begin{figure}[h]
%Fig.5
\includegraphics[width=8.5cm]{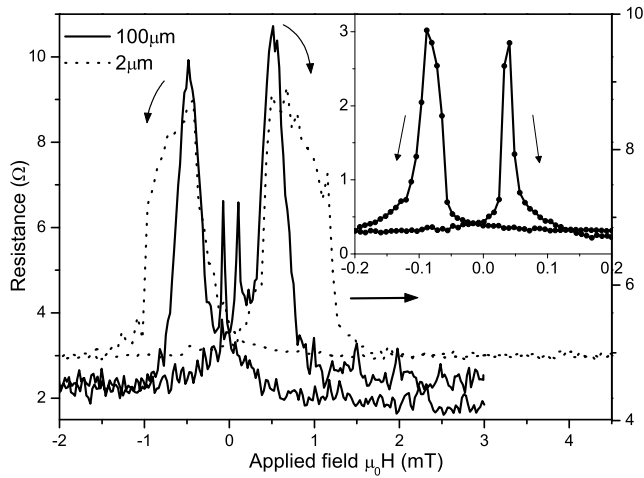}
\caption{\label{5mins_even}$R(H)$ curve for GdNi(11)/MoGe(42)/GdNi(11) trilayer at
$\sim 5$K for two different wire widths. Inset: GdNi(38)/MoGe(21)/GdNi(38) switching
below 0.1 mT at 2.53 K. Arrows denote the direction of the field sweep and the use
of the right-hand scale for the 2~$\mu$m structure.}
\end{figure}

The effect is not strongly sensitive to variation of the S and/or F layer
thicknesses. This is demonstrated by the data in Fig. \ref{5mins_even} in which MoGe
layer is thicker, as well as the inset of Fig. \ref{5mins_even} (thicker GdNi).
These all show qualitatively similar switching behavior to the original sample. In
the case of the thicker MoGe layer some additional features are observed around zero
field. These are the two F layers switching independently, (also confirmed by
magnetoresistance measurements above T$_c$ - not shown here). This is most likely a
combination of a reduction of the direct coupling between the F layers for thicker
MoGe, and an increase of the roughness and therefore $H_c$ of the top GdNi layer for
a thicker spacer. With thicker GdNi the $H_c$ is reduced, the peaks can shifted to
below 0.1 mT leading to a sensitivity in the switching at the steepest part of the
curve (increasing field sweep) $>60$$\Omega$/mT at an applied field of $\sim
90$$\mu$T. In Fig. \ref{5mins_even} we also show the effect of patterning wires of
different widths. The switching field changes from a 100$\mu$m to 2$\mu$m width due
to shape anisotropy, and with it therefore the field at which vortices are present
in the MoGe. This allows us further control over $H_c$ in complement to varying the
thickness of the GdNi.

\begin{figure}[ht]
%Fig. 6
\includegraphics[width=8.5cm]{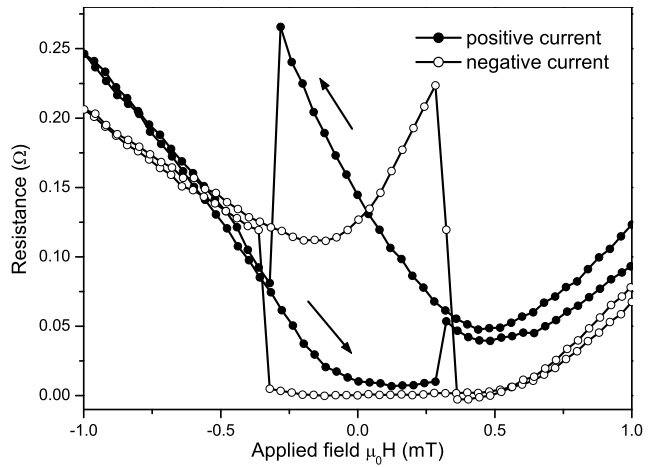}
\caption{\label{bilayerRHs}$R(H)$ for the bilayer MoGe(21)/GdNi(22) at 3.88 K, and
current directions as indicated; a positive current is defined as parallel to
positive $H$. The arrows denote the direction of the field sweep.}
\end{figure}

A point of interest is that in bilayer samples the switching behavior is less
pronounced and also more complicated, since we now find asymmetry in the peak value
of the voltage (resistance), but also asymmetry with respect to the current
direction. Fields sweeps are shown in Fig. \ref{bilayerRHs} for a bilayer sample
MoGe(21)/GdNi(22), at a temperature of 3.88~K, near the bottom of the transition.
For positive current (parallel to positive $H$) a jump rather than a peak is seen at
$+H_c$, and a rather broad peak occurs at $-H_c$. We also observe an increasingly
resistive background (suppression of the superconductivity) at higher fields. For
the other current direction the reverse is the case: a peak in $R$ occurs at $+H_c$
when sweeping from negative $H$. Unraveling this behavior would need extensive study
of the I-V characteristics, which will be for future work. We can, however, identify
several differences between trilayers and bilayers. For instance, the bilayer is in
fact asymmetric : the order parameter is strongly suppressed at the S/F interface,
but not at the free interface, and vortex pinning may actually be sensitive to
(inhomogeneities at) the free interface. Also, given the observation of only one
resistance peak in the trilayer case it appears that the domain walls couple across
the MoGe layer, at least for relatively thin MoGe. We made similar observations of
only one resistance peak for trilayers with two different thicknesses of the
F-layer, such as GdNi(11)/MoGe(21)/GdNi(22). We can speculate that this domain wall
coupling both enhances the local flux density in the superconductor and sharpens the
switching behavior.

\section{Vortex formation}
We have argued above that the magnetoresistive peaks we observe are due to vortices
formed in the S layer above Bloch walls present in the F layer. To make this more
compelling, we now estimate whether such vortices can actually be expected to form.
For this we use the results from the model discussed recently by Burmistrov and
Chtchelkatchev.\cite{burmistrov05} Basically, they find the amount of flux coming
out of a Bloch wall of width $\delta$ situated in a ferromagnetic film of thickness
$d_F$ characterized by a volume (saturation) magnetization $M_s$, and from a free
energy consideration calculate whether this can lead to the formation of a vortex in
a superconducting layer of thickness $d_S$, characterized by a London penetration
depth $\lambda_L$, placed on top of the ferromagnet. For the case of a single domain
wall they find, for given $\delta$, $d_F$, $d_S$ and $\lambda_L$, the minimum or
critical magnetization $M_c$ needed to overcome the lower critical field $H_{c1}$ of
the superconductor. Since in our experiments $M_s$ is a materials constant, it is
more convenient to write this condition in terms of a minimum thickness for the
F-layer $d_F^{min}$, which takes the form
\begin{equation}
d_F^{min} = \frac{\lambda H_{c1} }{M_s} \; \times \left\{
\begin{array}{ll}
        2 \lambda / \delta \; & \; (\pi \delta) \ll 4 \lambda \\
        1-32G \lambda / (\pi^2\delta) \; & \; (\pi \delta) \gg 4 \lambda
\end{array} \right.
\label{eq:dfmin}
\end{equation}
Here, $\lambda$ = $\lambda_L^2 / d_S$ if $d_S < \lambda_L$, otherwise $\lambda$
equals $\lambda_L$; $H_{c1}$ is given by $(\Phi_0 / 4 \pi \mu_0 \lambda^2)
\ln(\lambda / \xi)$, with $\xi$ the Ginzburg-Landau coherence length and $\Phi_0$
the flux quantum; G $\approx$ 0.916 is the Catalan constant, and $S.I.$ units. For
MoGe, the relevant values are $\lambda_L \approx$ 0.7~$\mu$m, which for a 20~nm film
yields $\lambda = $24.5~$\mu$m; and $\xi \approx$ 5~nm, leading to $H_{c1} \approx$
1.8~A/m, an extremely low value which is due to the combination of a large bulk
penetration depth and a small film thickness. For GdNi, the relevant characteristic
are $M_s$ and $\delta$. As discussed above, $M_s$ = 7.8$\times 10^5$~A/m
(corresponding to 0.98~T) is relatively large. Values for $\delta$ are not exactly
known, but the weak magnetocrystalline anisotropy leads to large wall widths, which
we take of order 1~$\mu$m. The relevant limit is then $\pi \delta \ll 4 \lambda$,
and eq.~\ref{eq:dfmin} yields $d_F^{min}$ =~2.9~nm. For the thicknesses we use, and
under the assumption of Bloch walls, the flux from a domain wall is therefore easily
large enough to create vortices. \\
This is the main conclusion from the calculation, but several more remarks are in
order. First, it is interesting to note that, in this limit, $d_F^{min}$ does not
depend on $\lambda$ (apart from the logarithmic factor $\ln(\kappa)$), since $H_{c1}
\propto 1/\lambda^2$. Then, we have disregarded the effect of the in-plane applied
field. In terms of the model, this is allowed since $d_S << \lambda$, which means
that the field fully penetrates without more than the vacuum contribution to the
free energy. Experimentally, it can be noted that misalignment effects are
apparently not relevant, since vortices are only created in significant amounts in
the domain state of the ferromagnet. Making a rough estimate, an applied field of
2~mT (outside the flux flow peak) under a 1$^{\circ}$ misalignment yields an
induction of $3 \times 10^{-2}$~A/m, again much smaller than the estimated $H_{c1}$.
Furthermore, we note that, although the flux density from the domain wall is well
above the lower critical field $\mu_0 H_{c1}$, it is still much lower than the upper
critical field $\mu_0 H_{c2}$. With a typical value of -$\mu_0 dH_{c2}/dT \approx$
2.5~T/K, and taking T-T$_c$ $\approx$ 100~mK, $\mu_0 H_{c2}(T)$ is found to be
$\approx$ 0.25~T, very much larger than $\mu_0 H_{c1}$. In summary therefore, the
reason that vortices can be observed close to the resistive transition in our
MoGe/GdNi system is that the amorphous ferromagnet combines a reasonably large
magnetization with a large domain wall width, while the amorphous superconductor
combines a large penetration depth with a high upper critical field.

Given these different constraints, it is of interest to consider the possibility of
vortex formation in various S/F systems which are currently under investigation as
$\pi$-junctions or spin switches, especially those based on Nb such as Nb/Py, Nb/Co
or Nb/CuNi. The F layers in these combinations are qualitatively different, with Py
having large $M_s$ and large $\delta$, Co having large $M_s$ and small $\delta$, and
CuNi having small $M_s$ and larger $\delta$. Also considered can be
YBa$_2$Cu$_3$O$_7$ (YBCO) and La$_{0.7}$Ca$_{0.3}$MnO$_3$ (LCMO). For Nb, we use
typical values of $d_{S}$ = 50~nm, $\lambda_L$ = 50~nm, $\xi$ = 12~nm, for YBCO we
use $d_{S}$ = 50~nm, $\lambda_L$ = 180~nm, $\xi$ = 2~nm. The different values for
the ferromagnets are given in Table~\ref{table:dfmin}, together with the computed
value for $d_F^{min}$. This is of course based on the assumption of Bloch and not
N{\'e}el walls in such thin films, which may not be the case, but the numbers are
instructive nonetheless.
\begin{table}[t]
\begin{tabular}{|c|c|c|c|c|} \hline
System          &   $\mu_0 M_s$ [T]   &  $\delta$ [$\mu$m]  &    $d_F^{min}$ [nm] &
($\pi \delta) / (4 \lambda$)
\\ \hline
a-MoGe/a-GdNi   &     0.98             &   1                 &   2.9    &   0.03        \\
Nb/Py           &     0.7             &   1                 &   5.7     &  16           \\
Nb/Co           &     1.6             &   0.05              &    5.9    &  0.8         \\
Nb/CuNi(50)    &     0.1             &    0.25              &    19.1  &  0.16             \\
Nb/CuNi(10)    &     0.1             &    0.5              &    20  &  1.6            \\
YBCO/LCMO       &     0.75             &    0.05              &    51   &  0.06
\\ \hline
\end{tabular}
\label{table:dfmin} \caption{Comparison of approximate critical thickness of F layer
to achieve vortices above a Bloch domain wall for various S and F materials. The
column entries are : the combination of materials, $\mu_0 M_s$, the estimate for
$\delta$, the calculated $d_F^{min}$ and the relevant limit for using
eq.~\ref{eq:dfmin}. Two cases are given for Nb/CuNi, one with the general S-layer
thickness of 50~nm, and one with $d_S$ = 10~nm, as used in Ref.~\cite{ryazanov03}.}
\end{table}
The table shows that the combination MoGe/GdNi actually yields the lowest value for
$d_F^{min}$ due to the combination strong magnet / large domain wall. Still, for the
strong magnetsthe numbers do not vary overmuch, $d_F^{min}$ is typically a few nm.
For weak CuNi it is significantly larger, which is interesting in view of the
observations of Ryazanov {\it et al.} \cite{ryazanov03}. They found flux flow
behavior in the $I-V$ curves of a Nb wire at in-plane applied fields around the
coercive field of a block of 18~nm thick CuNi on top of a portion of the wire and
ascribed the effect to vortices induced in the S layer due to Bloch domain walls in
the CuNi. Taking into account that the Nb layer in their case was only 10~nm thick,
the estimated value $d_F^{min}$ is 20~nm, which is roughly the thickness used in the
experiment. The Bloch wall scenario for this experiment appears not unreasonable,
since the prepared state is in-plane magnetized, while CuNi has a tendency to
perpendicular magnetization in this thickness range, as found for CuNi/Cu
multilayers \cite{ruotolo04}. The largest value for $d_F^{min}$ is found for
YBCO/LCMO, which is due to the large value of $\ln(\kappa)$ in this system.

\section{Conclusions}
We have demonstrated that amorphous F/S/F heterostructures can show large
magnetoresistance associated with vortex motion in the S layer, induced by magnetic
domains in the F layers. This magnetoresistance can be several tens of Ohms change
in a field step of a few tens of $\mu$T due to the combination of weak domain wall
and vortex pinning in these amorphous materials. We note that this effect can be a
relatively simple test for the presence of Bloch domain walls in a ferromagnetic
film. Also, the strong signals may provide a possibility to combine magnetic domain
and flux logic\cite{allwood02,reichhardt04} in a flexible way, since we have
demonstrated that both the GdNi thickness and wire widths can be effective tools to
tune the fields at which the peaks in flux flow resistivity are observed. That said,
we also should mention some problems open for further research. One point we have
not touched is the obvious question whether the measured increase in resistance can
be tied to flux-flow resistivity $\rho_{FF}$ in a quantitative way from the standard
formula $\rho_{FF} = \rho_n H/H_{c2}$, with $\rho_n$ the normal-state resistance. At
the moment we cannot answer that question since, apart from the fact that our
measurements have not been performed in the linear regime of the $I(V)$
characteristics where homogeneous flow can be assumed, we know neither the local
field, nor the amount of vortices (determined by the domain wall width) between the
voltage contacts. Also disregarded in the discussion are possible geometrical
effects which would lower the entry field for vortices due to an inhomogeneous
current distribution. This touches different questions such as whether this allows
smaller values than $d_F^{min}$, but also whether the nucleation of magnetic
domains, through their creation of vortices, actually is facilitated by edges or
corners. Fabrication of structures with artificial nucleation points would be an
interesting extension of the present work.

\begin{acknowledgments}
We thank B. Leerink for preliminary measurements on the GdNi system, and R. Hendrikx
and M. Hesselberth for X-ray and Rutherford Backscattering measurements. This work
is part of the research programme of the `Stichting voor Fundamenteel Onderzoek der
Materie (FOM)', which is financially supported by the `Nederlandse Organisatie voor
Wetenschappelijk Onderzoek (NWO)'.
\end{acknowledgments}


\begin{thebibliography}{99}
\bibitem{buzdin05} A.~I. Buzdin, Rev. Mod. Phys. {\bf 77}, 935 (2005).
%
\bibitem{tagirov99} L.~R. Tagirov, Phys. Rev. Lett.{\bf 83}, 2058 (1999).
%
\bibitem{baladie03} I. Baladi{\'e} and A. Buzdin, Phys. Rev. B {\bf 67}, 014523 (2003).
%
\bibitem{fominov03} Y.Fominov, A. A. Golubov and M.~Yu. Kupriyanov,
Pis'ma Zh. \'Eksp. Teor. Fiz. {\bf 77}, 609 (2003); [JETP Lett. {\bf 77}, 510
(2003)].
%
\bibitem{gu02} J.~Y.Gu, C.-Y. You, J.~S. Jiang, J. Pearson, Y.~B.Bazaliy and S. D. Bader,
Phys. Rev. Lett.{\bf 89}, 267001 (2002).
%
\bibitem{rusanov06} A.~Y.Rusanov, S. Habraken and J. Aarts, Phys. Rev. B {\bf 73},
060505(R) (2006).
%
\bibitem{kinsey01} R.~J. Kinsey, G. Burnell and M.~G. Blamire, IEEE Trans. Appl. Supercond. {\bf 11},
904 (2001).
%
\bibitem{yang04} Z. Yang, M. Lange, A. Volodin, R. Szymczak, and V.~V. Moshchalkov, Nature Materials
{\bf 3}, 793 (2004).
%
\bibitem{lange03} M. Lange, M. J. Van Bael, and V.~V.Moshchalkov, Phys. Rev. B {\bf 68}, 174522 (2003).
%
\bibitem{gillijns05} W. Gillijns, A. Yu. Aladyshkin, M. Lange, M. J. Van Bael and V.~V. Moshchalkov,
Phys. Rev. Lett. {\bf 95}, 227003 (2005).
%
\bibitem{ryazanov03} V.~V. Ryazanov, V.~A. Oboznov, A.~S. Prokof'ev and S.~V. Dubonos,
Pis'ma Zh. \'Eksp. Teor. Fiz. {\bf 77}, 43 (2003); [JETP Lett. {\bf 77} 39 (2003)].
%
\bibitem{burmistrov05} I.~S. Burmistrov, and N.~M. Chtchelkatchev, Phys. Rev. B {\bf
72}, 144520 (2005).
%
\bibitem{plourde02} B. L. T. Plourde, D. J. Van Harlingen, N. Saha, R. Besseling, M. B. S. Hesselberth and P.
H. Kes, Phys. Rev. B {\bf 66}, 054529 (2002).
%
\bibitem{baarle03} G. J. C. van Baarle, A.~M. Troianovski, T. Nishizaki, P.~H. Kes
and J. Aarts, Appl. Phys. Lett.{\bf 82}, 1081 (2003).
%
\bibitem{moorjani84} K. Moorjani and J. M. D. Coey, {\it Magnetic glasses}
(Elsevier, Amsterdam, 1984).
%
\bibitem{asomoza79} R. Asomoza, A. Fert, I.~A. Campbell, A. Li{\'e}nard and
J.~P. Rebouillat, J. Phys. F: Met. Phys. {\bf 9}, 349 (1979).
%
\bibitem{note-density} We use a density of 6.0$\times 10^{28}$ at/m$^3$, see A.
Gangulee and R. C. Taylor, J. Appl. Phys. {\bf 49}, 1762 (1978).
%
\bibitem{note-diamag} The magnetization even becomes slightly negative because of
the diamagnetic contribution of the Si substrate.
%
\bibitem{pena05} V. Pe{\~n}a, Z. Sefrioui, D. Arias, C. Leon, J. Santamaria, J.~L. Martinez,
S. G. E. teVelthuis and A. Hoffmann, Phys. Rev. Lett. {\bf 94}, 057002 (2005).
%
\bibitem{ruotolo04} A. Ruotolo, C. Bell, C. W. Leung, and M. G. Blamire, J. Appl.
Phys. {\bf 96}, 512 (2004).
%
\bibitem{allwood02} D.~A. Allwood, G. Xiong, M.~D. Cooke, C.~C. Faulkner, D. Atkinson, N. Vernier and
R.~P. Cowburn, Science{\bf 296}, 2003 (2002).
%
\bibitem{reichhardt04} C. J. Olsen Reichhardt, C. Reichhardt, M. J. Hastings and
B. Jank{\'o}, Physica C {\bf 404}, 266 (2004).

\end{thebibliography}
\end{document}